\documentclass[%
 jmp,
 amsmath,amssymb,
 reprint,%
]{revtex4-1}

\usepackage{graphicx}
\usepackage{dcolumn}
\usepackage{bm}

\usepackage[utf8]{inputenc}
\usepackage[T1]{fontenc}
\usepackage{mathptmx}
\newcommand{\beq}{\begin{equation}}
\newcommand{\eeq}{\end{equation}}
\newcommand{\bea}{\begin{eqnarray}}
\newcommand{\eea}{\end{eqnarray}}
\newcommand{\bwd}{\begin{widetext}}
\newcommand{\ewd}{\end{widetext}}

\def    \mrm  {\mathrm}

\begin{document}

\preprint{AIP/123-QED}

\title[Sample title]{On the gauge transformation for the rotation of the singular string  in the Dirac  monopole theory}

\author{Xiao-Yin Pan}
\affiliation{
Department of Physics, Ningbo University, Ningbo, Zhejiang 315211, China
}%

\author{Yin Chen}
\affiliation{
Department of Physics, Ningbo University, Ningbo, Zhejiang 315211, China
}%

\author{Yu-Qi Li}
\affiliation{
Shanghai Key Laboratory of Trustworthy Computing, East China Normal University, Shanghai 200062, China
}%

\author{Aaron G. Kogan}
\affiliation{Department of Physics, Massachusetts Institute of Technology, Cambridge, Massachusetts 02139, USA}

\author{Juhao Wu}
\affiliation{%
Stanford University, Stanford, California 94309, USA
}%

\date{\today}

\begin{abstract}
In the Dirac theory of the quantum-mechanical interaction of a magnetic monopole and an electric charge, the vector potential is singular from the origin to infinity along certain direction - the so called Dirac string. Imposing the famous quantization condition, the singular string attached to the monopole can be rotated arbitrarily by a gauge transformation, and hence is not physically observable. By deriving its analytical expression and analyzing its properties, we show that the gauge function $\chi({\bf r})$ which rotates the string to another one has quite complicated behaviors depending on which side from which the position variable ${\bf r}$ gets across the plane expanded by the two strings. Consequently, some misunderstandings in the literature are clarified.
\end{abstract}

\maketitle

\section{Introduction}
The magnetic monopole, though not been detected in nature so far, has become an important and fascinating research topic in many areas of physics \cite{1,2,3,4,5,6,7,8,9,kuhne97,10,11,12,13,Mavromatos20} since the seminal work of Dirac \cite{1,2}, who was seeking for an explanation of the observed fact that the electric charge is always quantized. The basis of this argument is that the vector potential Dirac introduced actually is a one for a magnetic monopole attached to an infinitely long and infinitesimally thin solenoid, the so called ``Dirac string''. In order to make the string unobservable, the Dirac quantization condition \cite{1} is required while assuming the wave function of the electric charge is single-valued. Thus, the existence of just one monopole anywhere in the universe would explain why the electric charge is quantized.  \\

Since the Dirac string is not observable, different positions of the string must give physically equivalent results. Actually the change of the string position is described by a gauge
transformation \cite{3,4,5,6,14,15}. This gauge transformation rotating the string from a position given by a unit vector to another new direction has a geometrical interpretation. However, its analytical expression is lack of and leads to considerable amount of misunderstandings and controversies \cite{7,16,17}, though non-singular vector potentials was used by Wu-Yang \cite{18,19} but at the expense of certain amount of topological complexity: space is divided into two overlapping regions in each of which the vector potential is continuous, and the potentials are related by gauge transformation in the overlap regions. The purpose of present work is to derive an analytical expression for the gauge function, and clarify some  misunderstandings in the literature.\\

In the following we will review briefly the quantum theory of a charge-monopole system in Sec. \ref{charge-monople}, then we derive the analytical expression for the gauge function, consequently its properties are analyzed in Sec. \ref{gaugefunction}. Discussion and concluding remarks are made in the last section \ref{discussion}.

\section{Quantum theory of the charge-monopole system}\label{charge-monople}
In 1931, Dirac first introduced the quantum mechanics of a magnetic monopole \cite{1}. He considered a system of an electron of electric charge $e$ and mass $m$ in the field a magnetic monopole sitting at the origin of the coordinate, with the magnetic field ${\bf B}_g=g\frac{{\bf r}}{r^3}$, where $g$ being the magnetic charge. Then the Hamiltonian for the system is
    \begin{eqnarray}
    \hat{H}_0({\bf r}, {\bf p})={1\over 2m}\left( {\hat {\bf p}}-{e\over c}{\bf A({\bf r})} \right)^2,
    \end{eqnarray}
where ${\bf r}, {\bf p}$ are the position and momentum operator of the electron, ${\bf A}({\bf r})$ being the vector potential of the magnetic monopole. Note that the electron is not allowed to pass through the monopole, or the wave function must vanish at the origin \cite{20,21} due to the fact that the Jacobi identity is not satisfied for $({\bf p}-\frac{e}{c}{\bf A})$. The Jacobi identity is recovered only if all of the species have the same ratio of electric to magnetic charge or if an electron and a monopole can never collide \cite{Heninger20}. Dirac proposed the following form of the vector potential ,
    \begin{eqnarray}\label{an}
    {\bf A}_{{\bf n} }({\bf r})&=& \frac {g}{r} \frac{{\bf r}\times {\hat n}}{ r-{\hat n}\cdot{\bf r}},
    \end{eqnarray}
where the unit vector ${\bf n}$ is directed along the $z$-axis: ${\bf n}=(0,0,1)$. This is the celebrated \emph{Dirac potential} \cite{1}. A straightforward calculation shows that the
magnetic field corresponds to this vector potential is
    \begin{eqnarray}
    {\bf B}({\bf r})&=& {\bf B}_g+{\bf B}_{string}={\bf B}_g-4\pi g {\bf n}\theta(z)\delta(x)\delta(y).
    \end{eqnarray}
On the other hand, it is shown \cite{22} the vector potential of Eq. (\ref{an}) can be rewritten as,
    \begin{eqnarray}\label{anint}
    {\bf A}_{{\bf n} }({\bf r})&=& g \int_{{\bf L}} \frac{({\bf r}-{\bf r'})\times d{\bf r'}}{|{\bf r}-{\bf r'}|^3},
    \end{eqnarray}
where the integral has to be taken along the straight line ${\bf L}$ which starts from the position of the monopole to infinity directed to the unit vector ${\bf n}$. Thus, this vector potential  actually corresponds not to a single isolated magnetic monopole, but rather to a straight solenoid of zero thickness from the origin to infinity along the direction ${\bf n}$ (the ``Dirac string'' $S_{\bf n}$), and the vector potential is singular along the Dirac string. In addition to the spherically symmetric Coulomb magnetic field, the appearance of an extra field of ${\bf B}_{string}$ directed along the singularity is unexpected.\\
\begin{figure}[htp]
  \centering
    \includegraphics[bb=0 0 470 470, angle=0, scale=0.5]{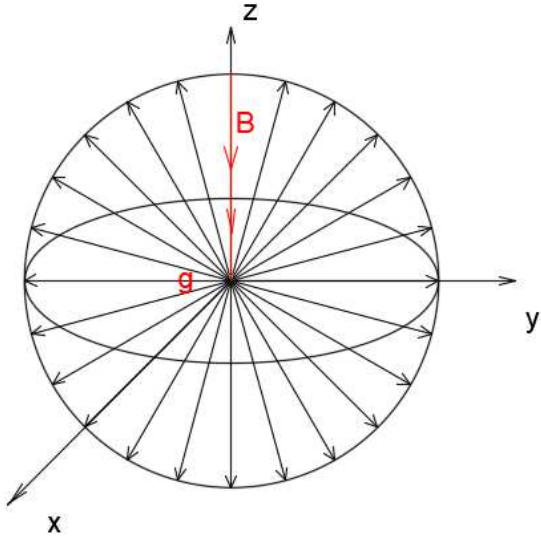}
    \caption{ (Color online). The magnetic field of the Dirac potential described by Eq.(3).}
\end{figure}

\begin{figure}[htp]
  \centering
    \includegraphics[bb=1 1 430 390, angle=0, scale=0.6]{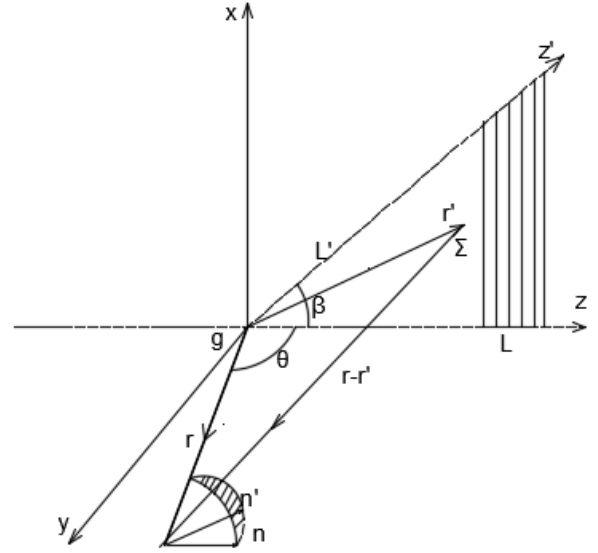}
    \caption{ The solid angle $\Omega_{{\bf nn'}}({\bf r})$ seen from position ${\bf r}$. The angle between axis $z$ and $z'$ is $\beta$.}
    \label{solidangle}
\end{figure}
In order to get rid of this problem, Dirac stated but did not prove \cite{1} a necessary and sufficient condition \cite{14} for the consistency of the theory, namely the remarkable quantization condition
    \begin{eqnarray}
    &&q=\frac{eg}{\hbar c}=\frac{n}{2}, (n=0,\pm1,\pm2,...).
    \end{eqnarray}
With this quantization condition, the (static part of the) quantum wave function of the electron in the static magnetic field of a monopole can be single valued. Later it was shown \cite{4,5,15} that this condition actually ensures that the position of the string is defined up to a gauge transformation, which is equivalent to rotating the line of singularity $S_{\bf n}$. In other words, the direction of the Dirac string is not fixed and could be arbitrary, therefore its field is not physical. To be specific, the gauge transformation which rotates the line of singularity from a position given by unit vector ${\bf n}$ to a new direction along another unit vector ${\bf n'}$ is
    \begin{eqnarray}\label{anminan}
    {\bf A}_{{\bf n'} }({\bf r})-{\bf A}_{{\bf n} }({\bf r})&=&\bigtriangledown \chi({\bf r}).
    \end{eqnarray}
By using Eq. (\ref{anint}) one obtains
    \begin{eqnarray}
    \bigtriangledown \chi({\bf r})&=& g \int_{{\bf C}} \frac{({\bf r}-{\bf r'})\times d{\bf r'}}{|{\bf r}-{\bf r'}|^3},
    \end{eqnarray}
where the integration is taken along the curve ${\bf C} = {\bf L'}-{\bf L}$. After some algebraic manipulations, one arrived at \cite{6,17,23},
    \begin{eqnarray}\label{delkitwo}
    \bigtriangledown \chi({\bf r}) &=& g \nabla \Omega_{{\bf n, n'}}({\bf r})-4\pi g\int_\Sigma \delta^{}({\bf r}-{\bf r'}) d{\bf S'},
    \end{eqnarray}
where
    \begin{eqnarray}
    \Omega_{{\bf n,n'} }({\bf r})&=&\int_\Sigma \frac{({\bf r}-{\bf r'})}{|{\bf r}-{\bf r'}|^3}\cdot  d{\bf S'}.
    \end{eqnarray}
It is clear that $\Omega_{{\bf n,n'} }({\bf r})$ is the solid angle under which the surface $\Sigma$ is seen from the point ${\bf r}$ (See Fig. \ref{solidangle}), with $\Sigma$ denotes the surface spanned by closing the curve ${\bf C}$ at infinity since the contribution of the infinite separated singular magnetic flux along this segment of ${\bf C}$ is vanishing.\\

The last term in Eq. (\ref{delkitwo}) is usually dropped \cite{7,24}, but this omission is misleading as pointed out in Ref. \cite{17}. It has also been understood \cite{3,6} that the function $\Omega_{{\bf nn'} }({\bf r})$ is discontinuous when ${\bf r}$ goes across the surface $\Sigma$, and the last term in Eq. (\ref{delkitwo}) cancels the corresponding $\delta$-function singularity, so that the result is continuous everywhere except on ${\bf L}$ and ${\bf L'}$. However, as we shall see below, depending on which side from which the position variable ${\bf r}$ gets across the plane expanded by the two strings, the last term in Eq. (\ref{delkitwo}) does not always cancel the corresponding $\delta$-function singularity. Hence, this statement is also incorrect.\\

\section{Calculation of the gauge function}\label{gaugefunction}
In order to investigate the properties of the gauge function $\chi({\bf r})$, next we proceed to calculate the analytical expression for the solid angle $\Omega_{{\bf nn'} }({\bf r})$. Without loss of generality, we can set the position of the magnetic charge as the original of our coordinates, choose the straight line ${\bf L }$ as the $z$-axis. The rotated line ${\bf L'}$ lies in the $x$-$z$ plane, the angle between ${\bf L }$ and ${\bf L' }$ is $\beta$, as depicted in Fig. \ref{solidangle}.\\

Working in the spherical coordinates, denoting ${\bf r} = r(\sin \theta \cos \varphi$, $\sin \theta \sin \varphi$, $\cos \theta)$, and noticing that ${\bf r'}$ lies in the $x$-$z$ plane, which can be written as ${\bf r'} = (r'\sin \theta'$, 0, $r'\cos\theta')$, then we have
    \begin{equation}\label{solidint}
    \Omega_{{\bf n,n'} }({\bf r}) =\int_\Sigma \frac {(-y)\cdot  r'dr' d\theta'}{|{\bf r} - {\bf r}'|^{3}}.\\
    \end{equation}
with the distance between ${\bf r}$ and  ${\bf r}'$,  $|{\bf r}-{\bf r}'| = [r^2+r'^2 - 2rr'(\sin \theta \cos \varphi \sin \theta ' + \cos \theta \cos \theta ')]^{1/2}$. It is evident that $\Omega_{{\bf nn'} }({\bf r})=0$ when $y=0$, namely when ${\bf r}$ lies in the plane spanned by the two strings. In the case when $y\neq 0$, using the following identity,
    \begin{equation}
    \int {\frac{{xdx}}{{\sqrt {{{ R}^3}} }}}  = \frac{{ - 2(2a + bx)}}{{\Delta \sqrt {R} }},
    \end{equation}
where ${R}(x) = a + bx + c{x^2}$ and $\Delta  = 4ac - {b^2}$, then performing the radial part of the integral in Eq. (\ref{solidint}), we finally arrive at,
    \begin{eqnarray}\label{omegannprev}
    &&\Omega_{{\bf n,n'} }({\bf r}) =\frac{y}{{ r}}\int_0^\beta  {} \frac{d\theta'}{[1 - (\cos \theta \cos \theta ' + \sin \theta \cos \varphi \sin \theta')]}
    \nonumber \\
    &&= \left\{
    \begin{array}{lll}
    2\left[\varphi-\frac{\pi}{2}-\arctan\left(\frac{\cot\frac{\theta}{2} \tan \frac{\beta}{2}-\cos\varphi}{\sin\varphi} \right)\right ], & \quad  y>0
    \\
    0, & y=0
    \\
    2\left[\varphi-\frac{3 \pi}{2}-\arctan\left(\frac{\cot\frac{\theta}{2} \tan \frac{\beta}{2}-\cos\varphi}{\sin\varphi} \right)\right ], & \quad y<0
    \end{array}
    \quad. \right.
    \nonumber \\
    \end{eqnarray}\\

\begin{figure}[htp]
  \centering
    \includegraphics[bb=10 10 600 430, angle=0, scale=0.42]{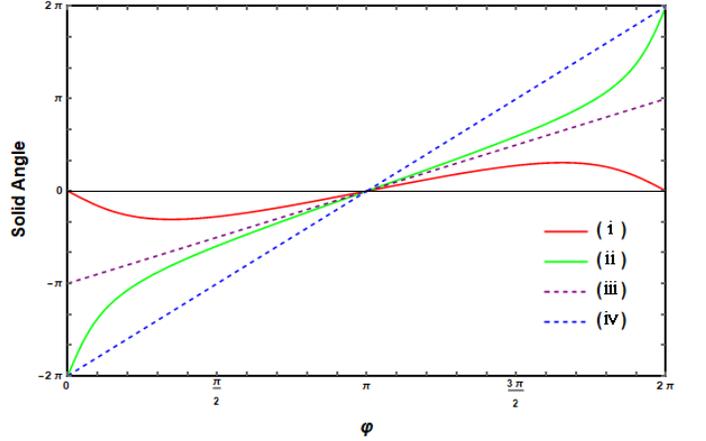}
    \caption{(Color online). The solid angle $\Omega_{{\bf nn'}}({\bf r})$ as functions of $\varphi$ at fixed values
    of $\theta$ and $\beta=\frac{\pi}{6}$ for cases
    (i)$\theta=\frac{\pi}{3} > \beta$;(ii) $\beta > \theta=\frac{\pi}{8}$  ;(iii) $\beta =\theta$ ; (iv)$\theta=0$. Note that the four curves are continuous at $\varphi=\pi$, but the values at $\varphi=0$ are different from those at $\varphi=2\pi$ for cases (ii),(iii) and (iv).}
    \label{solidcross}
\end{figure}

From Eq. (\ref{omegannprev}), it is evident that $\Omega_{{\bf n,n'} }({\bf r})$ is continuous at $\varphi=\pi$ or  $y=0$ when $x<0$, with value $\Omega_{{\bf n,n'} }({\bf r})|_{\varphi=\pi}=0$. This fact is also reflected in Fig. \ref{solidcross} where we have plotted the solid angle $\Omega_{{\bf nn'}}({\bf r})$ as functions of $\varphi$ at fixed values of $\theta$ and $\beta$.  It is evident that the four curves corresponding to the four cases listed below are all continuous at $\varphi=\pi$. However, it might be discontinuous when $\varphi$
crosses from $2\pi^-$ to $0^+$, {\it i.e.}, when $y$ crosses from $0^-$ to $0^+$ in the case when $x>0$. To make it clearer, we need to consider the following two cases: (i) $\theta > \beta $, then it is not difficult to obtain that $\Omega_{{\bf n,n'} }({\bf r})|_{x>0,y\rightarrow 0^+} = \Omega_{{\bf n,n'} }({\bf r})|_{\varphi\rightarrow 0^+} = 0$ and $\Omega_{{\bf n,n'} }({\bf r})|_{x>0, y \rightarrow 0^-} = \Omega_{{\bf n,n'} }({\bf r})|_{\varphi\rightarrow 2\pi^-} = 0$, thus $\Omega_{{\bf n,n'} }({\bf r})$ is continuous at $y=0$ when $x>0$; (ii) $\beta > \theta$, then $\Omega_{{\bf n,n'} }({\bf r})|_{x>0, y\rightarrow 0^+} = \Omega_{{\bf n,n'} }({\bf r})|_{\varphi\rightarrow 0^+} = - 2 \pi$ and $\Omega_{{\bf n,n'} }({\bf r})|_{x>0, y \rightarrow 0^-} = \Omega_{{\bf n,n'} }({\bf r})|_{\varphi\rightarrow 2\pi^-} = 2 \pi$, thus we have $\frac{\partial \Omega_{{\bf n,n'} }({\bf r})}{\partial y}|_{x>0,y=0} = - 4 \pi \delta(y)$.\\

In the literature \cite{3,6} the two strings ${\bf L}$ and ${\bf L' }$ were always excluded when  ${\bf r}$  as the variable of the gauge function $\chi({\bf r})$, goes across the surface $\Sigma$. But with the analytical expression at hand, it is not difficult to note that we shall consider the scenario when ${\bf r}$ gets across the surface through the two strings, namely the following two cases: (iii) $\beta = \theta$, then $\Omega_{{\bf n,n'} }({\bf r})|_{x>0, y\rightarrow 0^+} = \Omega_{{\bf n,n'} }({\bf r})|_{\varphi\rightarrow 0^+} = - \pi$, and $\Omega_{{\bf n,n'} }({\bf r})|_{x>0, y \rightarrow 0^-} = \Omega_{{\bf n,n'} }({\bf r})|_{\varphi\rightarrow 2\pi^-} = \pi$, thus $\frac{\partial \Omega_{{\bf n,n'} }({\bf r})}{\partial y}|_{x>0,y=0} = - 2 \pi \delta(y)$; (iv) $\theta = 0$, and $\Omega_{{\bf n,n'} }({\bf r})|_{x>0,y\rightarrow 0^+} = \Omega_{{\bf n,n'} }({\bf r})|_{\varphi\rightarrow 0^+} = - 2 \pi$, $\Omega_{{\bf n,n'} }({\bf r})|_{x>0, y \rightarrow 0^-} = \Omega_{{\bf n,n'} }({\bf r})|_{\varphi\rightarrow 2 \pi^-} = 2 \pi$, thus $\frac{\partial \Omega_{{\bf n,n'} }({\bf r})}{\partial y}|_{x>0,y=0} = - 4 \pi \delta(y)$. Hence, $\Omega_{{\bf n,n'} }({\bf r})$ is discontinuous when $\varphi$ crosses from $2\pi^-$ to $0^+$ for case (ii) $\beta >\theta$, (iii) $\beta =\theta$, and (iv) $\theta=0$. But surprisingly, the magnitudes of discontinuous for each case are different, this fact is also reflected in Fig. \ref{solidcross}.\\

For the special case when $\beta=\pi$, the solid angle $\Omega_{{\bf n, -n} }({\bf r})$ is ill-defined \cite{15}. However, if we define it as the limit of $\beta\rightarrow \pi$ rotating from $z$ axis in the $x-z$ plane, then from Eq. (\ref{omegannprev}) we have $\Omega_{{\bf n,n'} }({\bf r})=2(\varphi-\pi)$, for $ 0\leq\varphi<2\pi$. This is consistent with the result obtained in Ref. \cite{15}, and from which the the gauge function $2g\varphi$ that relates the vector potential in the overlap regions in the Wu-Yang monopole theory \cite{19} can also be obtained readily.\\

The combinations of Eq. (\ref{delkitwo}) and Eq. (\ref{omegannprev}) yields the expression for the gauge function,
    \begin{eqnarray}\label{kitwo}
    &&\chi({\bf r})= \Omega_{{\bf n,n'} }({\bf r})+\chi_1({\bf r}),
    \end{eqnarray}
where
    \begin{eqnarray}
    \chi_1({\bf r})
    &=&
    \left\{
    \begin{array}{ll}
    4\pi g \theta(y) &  {\mrm{if}} \quad \beta \geq \theta
    \\
    0 & {\mrm{if}} \quad \beta <\theta
    \end{array} \right.\quad ,
    \end{eqnarray}
whose gradient will give rise to the second term on the r.h.s of Eq. (\ref{delkitwo}), and $\theta(y) = \left\{ \begin{array}{ll}1 & {\mrm{if}}\quad  y>0 \\ 0 & {\mrm{if}}\quad y<0
\end{array}\quad\right.$, is the Heaviside function \cite{25} and whose derivative gives rise to the delta function, {\it i.e.},$\frac{d \theta(y)}{dy} = \delta(y)$. Therefore, when (a) $x>0, y=0$, we need to consider the following situations: for case (i) $\theta>\beta$, $\chi({\bf r})|_{x>0}$ is continuous at $y=0$; case (ii) $\beta>\theta$, $\chi({\bf r})|_{x>0,y\rightarrow 0^+}=2g\pi$, $\chi({\bf r})|_{x>0,y\rightarrow 0^-} = 2 g \pi$, thus $\chi({\bf r})|_{x>0}$ is continuous at $y=0$; (iii) $\beta=\theta$ and $\chi({\bf r})|_{x>0,y\rightarrow 0^+} = 3 g \pi$, $\chi({\bf r})|_{x>0,y\rightarrow 0^-} = g \pi$, thus $\frac{\partial \chi({\bf r})}{\partial y}|_{x>0,y=0} = 2 \pi g \delta(y)$; (iv) $\theta=0$,  $\chi({\bf r})|_{x>0,y\rightarrow 0^+} = 2 g \pi$, $\chi({\bf r})|_{x>0,y\rightarrow 0^-} = 2 g \pi$, thus $\chi({\bf r})|_{x>0}$ is continuous at $y=0$. In the case when $\beta>\theta$, it is apparent that $\chi_1({\bf r})$, the last term in Eq. (\ref{delkitwo}) cancels the $\delta$-function singularity as previously understood \cite{3,6}. However, this is not true when $\beta=\theta$. Actually, in this case the l.h.s of Eq. (\ref{anminan}) is singular, as such $\chi({\bf r})$ was also thought to be singular or not well defined \cite{14}.\\

For convenience, we summarize the values for  $\Omega_{{\bf n,n'}}$, $\chi_1({\bf r})$, and $\chi({\bf r}) $ in the limits of $\varphi \rightarrow{\pi^{\pm}}$ in Table \ref{tabresults}, and in the limits of $\varphi \rightarrow 0^+, 2\pi^-$ in Table \ref{tabresends}.  As clearly seen from  Table \ref{tabresults},   $\chi_1({\bf r})$ does not cancel, but adds the $\delta$-function singularity in all the four cases. Moreover, from Table \ref{tabresends} it is clear that $\chi_1({\bf r})$ does not play a role in case (i). $\chi_1({\bf r})$ cancels the
$\delta$-function singularity in cases (ii) and (iv), while adding the $\delta$-function singularity in case (iii). Hence, depending on from which side ${\bf r}$ gets across the plane expanded by the two strings, the gauge function has quite complicated behaviors since $\nabla\chi_1({\bf r})$ does not always cancel the $\delta$-function singularity. \\

\begin{table}[h]
    \centering
    \begin{tabular}{|p{0.9cm}|p{2.4cm}|p{2.4cm}|p{2.3cm}|c|c|c|c|}\hline
        case &\multicolumn{3}{c|}{${\rm{x < 0,y \rightarrow 0   }}(\varphi \rightarrow{\pi^{\pm}})$}\\\hline
        $\theta  > \beta $&$
        \Omega_{{\bf n,n'}}|_{\varphi  \rightarrow {\pi^{\pm}}} = 0 $&$
        {\chi_1}|_{\varphi\rightarrow{\pi^{\pm}}}
        = 0$
        &$ {\chi}{|_{_{\varphi  \rightarrow {\pi^{\pm} }}}}
        = 0
        $
        \\ \hline
        $\theta  < \beta $&$
        \Omega_{{\bf n,n'}} {|_{\varphi  \rightarrow {\pi^{\pm} }}} =
        0$&$\begin{array}{l} {\chi_1}|_{\varphi\rightarrow{\pi^-}}
        = 4\pi g\\
        {\chi_1}|_{\varphi \rightarrow{\pi^+ }}
        = 0
        \end{array}$&$\begin{array}{l}
        {\chi}{|_{_{\varphi  \rightarrow {\pi ^ - }}}}
        = 4\pi g\\
        {\chi}{|_{_{\varphi \rightarrow{\pi ^ + }}}}
        = 0
        \end{array}$\\ \hline \hline
        $\theta  =\beta$ & $
        \Omega_{{\bf n,n'}} {|_{\varphi  \rightarrow {\pi^{\pm} }}} = 0$
        &$\begin{array}{l} {\chi_1}|_{\varphi\rightarrow{\pi^-}}
        = 4\pi g\\
        {\chi_1}|_{\varphi \rightarrow{\pi^+ }}
        = 0
        \end{array}$ &$\begin{array}{l}
        {\chi}{|_{_{\varphi  \rightarrow {\pi ^ - }}}}
        = 4\pi g\\
        {\chi}{|_{_{\varphi \rightarrow{\pi ^ + }}}}
        = 0
        \end{array}$  \\ \hline
        $ \theta=0$ &$
        \Omega_{{\bf n,n'}} {|_{\varphi  \rightarrow {\pi^{\pm} }}} = 0$   &
        $\begin{array}{l} {\chi_1}|_{\varphi\rightarrow{\pi^-}}
        = 4\pi g\\
        {\chi_1}|_{\varphi \rightarrow{\pi^+ }}
        = 0
        \end{array} $  & $\begin{array}{l}
        {\chi}{|_{_{\varphi  \rightarrow {\pi ^ - }}}}
        = 4\pi g\\
        {\chi}{|_{_{\varphi \rightarrow{\pi ^ + }}}}
        = 0
        \end{array}$  \\ \hline
    \end{tabular}
    \caption{ Values for $\Omega_{{\bf n,n'}}$, $\chi_1({\bf r})$, and $\chi({\bf r}) $, in the limits of $\varphi \rightarrow{\pi^{\pm}}$.}
    \label{tabresults}
\end{table}

\begin{table}[h]
    \centering
    \begin{tabular}{|p{0.9cm}|p{2.7cm}|p{2.3cm}|p{2.1cm}|c|c|c|c|}\hline
        case &\multicolumn{3}{c|}{${\rm{x > 0,y \rightarrow 0   }}(\varphi \rightarrow{0^{+}, 2\pi^-})$}\\\hline
        $\theta > \beta $&$\begin{array}{l}
        \Omega_{{\bf n,n'}} {|_{\varphi  \rightarrow {0^ + }}} = 0\\
        \Omega_{{\bf n,n'}} {|_{\varphi  \rightarrow{2\pi^-}}} = 0
        \end{array}$&$\begin{array}{l}
        {\chi_1}|_{\varphi\rightarrow{0^+}}
        = 0\\
        {\chi_1}|_{\varphi \rightarrow{2\pi^- }}
        = 0
        \end{array}$
        &$\begin{array}{l} {\chi}{|_{_{\varphi  \rightarrow {0^+ }}}}
        = 0\\
        {\chi}{|_{_{\varphi \rightarrow{2\pi^- }}}}
        = 0
        \end{array}$
        \\ \hline
        $\theta  < \beta $&$\begin{array}{l}
        \Omega_{{\bf n,n'}} {|_{\varphi \rightarrow{0^ + }}} =-2\pi\\
        \Omega_{{\bf n,n'}} {|_{\varphi \rightarrow{2\pi^-}}} = 2\pi
        \end{array}$&$\begin{array}{l}
        {\chi_1}|_{\varphi\rightarrow{0^+}}
        = 4\pi g\\
        {\chi_1}|_{\varphi \rightarrow{2\pi^- }}
        = 0
        \end{array}$&$\begin{array}{l}
        {\chi}{|_{_{\varphi  \rightarrow {0^+ }}}}
        = 2\pi g\\
        {\chi}{|_{_{\varphi \rightarrow{2\pi^- }}}}
        = 2\pi g
        \end{array}$\\ \hline \hline
        $\theta  =\beta$ & $\begin{array}{l}
        \Omega_{{\bf n,n'}} {|_{\varphi  \rightarrow {0^+ }}} = -\pi\\
        \Omega_{{\bf n,n'}} {|_{\varphi  \rightarrow{2\pi^- }}} = \pi
        \end{array}$   &$\begin{array}{l}
        {\chi_1}|_{\varphi\rightarrow{0^+}}
        = 4\pi g\\
        {\chi_1}|_{\varphi \rightarrow{2\pi^- }}
        = 0
        \end{array}$ &$\begin{array}{l}
        {\chi}{|_{_{\varphi  \rightarrow {0^+ }}}}
        = 3\pi g\\
        {\chi}{|_{_{\varphi \rightarrow{2\pi^-}}}}
        = \pi g
        \end{array}$  \\ \hline
        $ \theta=0$ &$\begin{array}{l}
        \Omega_{{\bf n,n'}} {|_{\varphi  \rightarrow {0^+ }}} = -2\pi\\
        \Omega_{{\bf n,n'}} {|_{\varphi  \rightarrow{2\pi^-}}} = 2\pi
        \end{array}$   & $\begin{array}{l}
        {\chi_1}|_{\varphi\rightarrow{0^+}}
        = 4\pi g\\
        {\chi_1}|_{\varphi \rightarrow{2\pi^- }}
        = 0
        \end{array}$   & $\begin{array}{l}
        {\chi}{|_{_{\varphi  \rightarrow {0 ^ - }}}}
        = 2\pi g\\
        {\chi}{|_{_{\varphi \rightarrow{2\pi ^ - }}}}
        = 2\pi g
        \end{array}$  \\ \hline
    \end{tabular}
    \caption{ Values for $\Omega_{{\bf n,n'}}$, $\chi_1({\bf r})$, and $\chi({\bf r}) $, in the limits of $\varphi \rightarrow 0^+, 2\pi^-$.}
    \label{tabresends}
\end{table}

\section{Discussion and Conclusions}\label{discussion}
As a side issue, we next prove an identity which is related to the key result of Eq. (\ref{kitwo}) we obtained in the last section. It is clear that our process of calculations are also valid when the system is confined in some external central potential $V({r})$ of spherical symmetry, {\it i.e.}, when the Hamiltonian with string ${\bf n}$ is
    \begin{equation}
        {{\hat H}_{\bf n}} = \frac{1}{{2m}}{\left({\hat {\bf p}} - \frac{e}{c}{{\bf A}_{\bf n}}({\bf r})\right)^2} + V({r}),
    \end{equation}
and whose eigenfunction can be written as
    \begin{equation}
        \Psi_{\bf n}({\bf r})  =\mathcal{N} R_{n_r,l}(r){Y_{q,l,m}}(\theta,\phi ),
    \end{equation}
where $\mathcal{N}$ is the normalization constant, the angular part ${Y_{q,l,m}}(\theta ,\phi )$ are the monopole harmonics \cite{19}, and $R_{n_r,l}(r)$ are the radial part of the eigenfunction, satisfying
    \bea
    && \left[ - \frac{1}{{2m{r^2}}}\frac{\partial }{{\partial r}}\left({r^2}\frac{\partial }{{\partial r}}\right) + \frac{{l(l + 1) - {q^2}}}{{2m{r^2}}} + V(r)\right]R_{n_r,l}(r) 
    \nonumber \\
    &=& ER_{n_r,l}(r).
    \eea
The Hamiltonian with string ${\bf n'}$ is
    \begin{equation}
        {H_{\bf n'}} = \frac{1}{{2m}}{\left({\hat {\bf p}} - \frac{e}{c}{{\bf A}_{\bf n'}}({\bf r})\right)^2} + V({r}),
    \end{equation}
which is related to ${{\hat H}_{\bf n}}$ by the gauge transformation of Eq. (\ref{anminan}), {\it i.e.},
    \begin{equation}
        {{\hat H}_{\bf n'}} = U {{\hat H}_{\bf n}} U^{-1},
    \end{equation}
where  $U = {e^{\frac{{ - ie\chi ({\bf r})}}{{\hbar c}}}}$, and ${\Psi _{{\bf n'}}}({\bf r})$ as the eigenfunction of ${H_{\bf n'}}$ is related to $\Psi_{\bf n}({\bf r})$ as,
    \begin{equation}
        {\Psi _{{\bf n'}}}({\bf r})=U \Psi_{\bf n}({\bf r})= {e^{\frac{{ - ie\chi ({\bf r})}}{{\hbar c}}}}{\Psi _{\bf n}}({\bf r}).
    \end{equation}\\

It is apparent that the eigenfunction is invariant under a coordinate rotation ${\bf r}\rightarrow {\bf r'} = R {\bf r}$ and a simultaneous rotation of the string ${\bf n}\rightarrow {\bf n'} = R {\bf n}$, while the latter rotation can be undone by the gauge transformation ${\bf A}_{\bf n}\rightarrow {\bf A}_{\bf n'}$. Thus
we have
    \begin{eqnarray}
        &&\langle R^{-1}{\bf r}|{\Psi _{\bf n}}\rangle=\langle {\bf r}|{{\mathcal D}}(R) |\Psi _{\bf n'}\rangle,
    \end{eqnarray}
in other words,
    \begin{equation}
        {\Psi _{\bf n}}(R^{-1}{\bf r}) ={\hat {\mathcal D}}(R) {\Psi _{\bf n'}}({\bf r}) = {e^{\frac{{ - ie\chi ({\bf r})}}{{\hbar c}}}}{\hat {\mathcal D}}(R){\Psi _{\bf n}}({\bf r}).
    \end{equation}
Furthermore, since for fixed $q$, ${Y_{q,l,m}}(\theta ,\phi )$ form a complete set, {\it i.e.} $\sum_{l',m'}$ $\langle |ql'm'\rangle$ $\langle ql'm'| = 1$,
    \bea\label{dpsi}
    {\hat {\mathcal D}}(R){\Psi _{\bf n}}({\bf r})
    &=&\langle {\bf r}|{\mathcal D}(R)|\Psi_{\bf n}\rangle
    \nonumber \\
    &=&
    \sum_{l',m'}\langle {\bf r}|ql'm'\rangle \langle ql'm'| {\mathcal D}(R)|\Psi_{\bf n}\rangle.
    \eea
In general, ${\mathcal D}(R)$ can be expressed in terms of Euler angles as  ${\mathcal D}(R) = e^{i\alpha J_z}$ $e^{i\beta J_y}$ $e^{i\gamma J_z}$. In the special choice of the coordinates  as shown in Fig. \ref{solidangle}, ${\mathcal D}(R)=e^{i\beta J_y}$ denoting the rotation around $y$-axis counterclockwisely by angle $\beta$ with $J_y$ the $y$-component of the angular momentum ${\bf J} = {\bf r} \times ({\bf p}-e {A})-q \frac{{\bf r}}{r}$.  \\

By setting $|\Psi_{\bf n}\rangle = |qlm\rangle$, and note that
    \begin{eqnarray}\label{jxyz}
        (J_x\pm i J_y)|qlm\rangle&=&\sqrt{(l\mp m)(l\pm m+1)} |qlm\pm1\rangle, \nonumber\\ J_z|qlm\rangle&=&m |qlm\pm1\rangle,
    \end{eqnarray}
one immediately realizes that angular momentum operators $J_i,i=x,y,z$ can only  change the value $m$ but not $l$ when  operating on $|qlm\rangle$. Thus with the use of the orthonormality relationship $\langle ql'm'|qlm\rangle = \delta_{ll'}\delta_{mm'}$ \cite{27}, and since they will not affect the radial part of the eigenfunction, thus Eq. (\ref{dpsi}) becomes
    \begin{equation}
        {\hat {\mathcal D}}(R){Y_{qlm}(\theta,\phi)}=\sum_{m'}\langle {\bf r}|qlm'\rangle \langle qlm'| {\mathcal D}(R)|qlm\rangle.
    \end{equation}

Since the gauge function is merely dependent on the angle $(\theta,\varphi)$, we immediately have the identity,
    \begin{equation}
        {Y_{qlm}}(\theta-\beta,\phi )={e^{ - iq \Omega}}\sum_{m'}{Y_{qlm'}}(\theta ,\phi )d^l_{m'm}(\beta ),
    \end{equation}
where $\Omega = \Omega_{{\bf n,n'}}(\theta,\varphi)$, and the matrix $d^l_{m'm}(\beta)$ = $\langle qlm' | e^{i\beta J_y} | qlm \rangle$ = $\langle lm'| e^{i\beta J_y} | lm \rangle$. This equality can be readily proved by Taylor expanding $e^{i\beta J_y} = 1 + i \beta J_y + ...,$ and with the help of Eq. (\ref{jxyz}), it is evident that each term is independent of $q$. Here we have dropped $\chi_1({\bf r})$ since ${e^{\frac{{ - ie\chi_1 ({\bf r})}}{{\hbar c}}}} = 1$. Equation (\ref{jxyz}) is a special case of the additional theorem obtained in Ref. \cite{18, 28} due to the simpler choice of the coordinates we have made.\\

It is worth stressing two points. First, that the gauge function is discontinuous at the surface expanded by the two related strings, the vector potentials are expected to be smooth and well-behaved functions of ${\bf r}$ everywhere in space-except their respective strings. However, the gauge function is finite but discontinuous at those two strings as we discussed for cases (iii) and (iv). Second, the discontinuity $\Delta \chi = 2 \pi g, 4 \pi g$ in the gauge function makes itself multi-valued,  but $e^{-i\frac{e \chi}{\hbar c}}$ is single-valued since the discontinuity will only multiply the wave function by $e^{-i\frac{e \Delta \chi}{\hbar c}} = \pm 1$.
\\

In summary, we have obtained the analytical expression for the gauge function of the gauge transformation which rotates the singular string in the Dirac  monopole theory. It is shown that the gauge function $\chi({\bf r})$ is a multi-valued function, which has quite complicated behaviors at surface expanded by the two related strings. We expect our results can shed lights in application of the monopole theory.\\


\medskip



\end{document}